\documentclass[runningheads]{svmult}

\usepackage{makeidx}   % allows index generation
\usepackage{graphicx}  % standard LaTeX graphics tool
\usepackage{subeqnar}  % subnumbers individual equations
\usepackage{multicol}  % used for the two-column index
\usepackage{physprbb}  % modified textarea for proceedings,
\makeindex             % used for the subject index
\begin{document}
\title*{The Metallicity dependence of \protect\newline the Cepheid Period-Luminosity relation: \protect\newline methodology and results}
%
%
%\toctitle{Focusing of a Parallel Beam to Form a Point
%\protect\newline in the Particle Deflection Plane}
% allows explicit linebreak for the table of content
%
%
\titlerunning{The Metallicity dependence of the Cepheid P-L relation}
% allows abbreviation of title, if the full title is too long
% to fit in the running head
%
\author{Marta Mottini\inst{1}
\and Martino Romaniello\inst{1}
\and Francesca Primas\inst{1}
\and Martin Groenewegen\inst{2}
\and Giuseppe Bono\inst{3}
\and Patrick Fran\c{c}ois \inst{4}}
\authorrunning{Marta Mottini et al.}
% if there are more than two authors,
% please abbreviate author list for running head
%
%
\institute{European Southern Observatory, Karl-Schwarzschild Strasse 2, D-85748 Garching b. M\"{u}nchen, Germany
\and Istituut voor Sterrenkunde, Celestijnenlaan 200B, B-3001 Leuven, Belgium
\and INAF-Osservatorio Astronomico di Roma, via Frascati 33, I-00040 Monte Porzio Catone, Italy
\and Observatoire de Paris-Meudon, GEPI, 61 avenue de l'Observatoire, F-75014 Paris, France}

\maketitle              % typesets the title of the contribution

\begin{abstract}
We present the results of an observational campaign undertaken to assess the influence of the iron content on the Cepheid Period-Luminosity relation. Our data indicate that this dependence is not well represented by a simple linear relation. Rather, the behaviour is markedly non monotonic, with the correction peaking at about solar metallicity and declining for higher and lower values of $[Fe/H]$.
\end{abstract}

\section{Data analysis and results}
Cepheid stars, through their Period-Luminosity relation, are one of the pillars on which the extragalactic distance scale is built. To this day, however, the debate is still open on the role played by the chemical composition on the pulsational properties of these stars, with different theoretical models and observational results leading to markedly different conclusion (e.g. \cite{journ4}, \cite{journ8}, \cite{journ10}, \cite{journ11}). To tackle this problem we used high resolution spectra collected with UVES and FEROS .Our sample includes a total of 76 stars: 40 Galactic, 22 LMC, and 14 SMC Cepheids. 
As first step in our analysis of the chemical composition of Cepheids we focused on the iron content. We developed a robust analysis procedure in order to accurately determine the iron abundance. First, we carefully assembled a reliable list of 263 iron lines (both neutral and ionized) between 480 and 780\,nm. We selected the oscillator strenght from the VALD database and we visually inspected each line profile on the observed spectra in order to detect and eliminate those lines affected by other elemental blends. 
Secondly, we measured the equivalent widths of all the lines assembled as described above using the {\em FITLINE} routine based on Gaussian fit. Finally, our iron abundances were determined using Kurucz's {\em ATLAS9} model atmospheres and the {\em WIDTH9} code (\cite{journ13}).

%\section{Results}
We have determined the iron abundances for a sub-sample of stars: 13 Galactic, 13 LMC and 12 SMC Cepheids. 
Our main result is summarised in the Fig.1, where we plot the V-band residuals $\delta(M_{V})$ of our stars from the standard PL relation of \cite{journ9} as a function of the iron abundance we have derived from the FEROS and UVES spectra. This standard PL relation was derived for the LMC as a whole ($<[Fe/H]>\cong -0.4$) and $\delta(M_{V})$ is, effectively, the correction to be applied to a ``universal'' PL relation as a function of metallicity.
A positive $\delta(M_{V})$ means fainter than the standard PL relation.

\begin{figure}[t]
\begin{center}
\includegraphics[width=.5\textwidth]{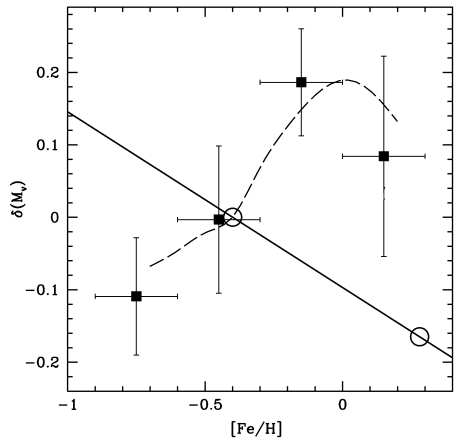}
\end{center}
\caption[]{Result from the subset analysed so far. The data are binned in metallicity to reflect the typical uncertainty on our determination of [Fe/H], marked by horizontal error-bars. The median value of $\delta(M_{V})$ in each metallicity bin is plotted as filled squares, with the vertical error-bars representing its associated error. For comparison we also plot the empirical results from \cite{journ11} in two Cepheid fields in M101 ({\em{open circles and solid line}}) and the theoretical predictions by \cite{journ8} from non-linear pulsational models (\em{dashed line})}
%\label{eps1}
\end{figure}

Our data indicate that the stars become fainter as metallicity increases, until a plateau or turnover point is reached at about solar metallicity. Our data are incompatible with both no dependence of th PL relation on iron abundance and with the linearly decreasing behaviour often found in the literature (e.g. \cite{journ11}, \cite{journ17}). On the other hand, non-linear theoretical models of \cite{journ8} provide a fairly good description of the data. For an in-depth discussion see \cite{journ15}.

%INDEX%%%%%%%%%%%%%%%%%%%%%%%%%%%%%%%%%%%%%%%%%%%%%%%%%%%%%%%%%%%%%%%
% Please check with the editor of your book whether he plans to
% include a "mutual" subject index - if so, please code your entries
% in the standard syntax. For your own purposes you may print your
% "personal" index by using the following commands:
%
%\clearpage
%\addcontentsline{toc}{section}{Index}
%\flushbottom
%\printindex
%%%%%%%%%%%%%%%%%%%%%%%%%%%%%%%%%%%%%%%%%%%%%%%%%%%%%%%%%%%%%%%%%%%%%

\end{document}